\documentclass{article}

\usepackage[preprint]{myStyle_2019}
\usepackage[utf8]{inputenc} 
\usepackage[T1]{fontenc}    
\usepackage{hyperref, color}       
\usepackage{url}            
\usepackage{booktabs,enumitem}       
\usepackage{amsfonts,amsmath,amssymb}       
\usepackage{nicefrac}       
\usepackage{microtype}      
\usepackage{graphicx}       
\newcommand*{\Scale}[2][4]{\scalebox{#1}{$#2$}}

\title{A Conditional Generative Model for Predicting Material Microstructures from Processing Methods}

\author{
  Akshay Iyer\thanks{The work was done during this author's internship at Siemens Corporate Technology.} \\
  Northwestern University\\
  Evanston, IL 60208 \\
  \texttt{akshayiyer2021@u.northwestern.edu} \\
  \And
  Biswadip Dey\\
  Siemens Corporate Technology\\
  Princeton, NJ 08536 \\
  \texttt{biswadip.dey@siemens.com}\\
  \AND
  Arindam Dasgupta\\
  Siemens Corporate Technology\\
  Princeton, NJ 08536 \\
  \texttt{arindam.dasgupta@siemens.com}\\
  \And
  Wei Chen\\
  Northwestern University\\
  Evanston, IL 60208 \\
  \texttt{weichen@northwestern.edu}\\
  \AND
  Amit Chakraborty\\
  Siemens Corporate Technology\\
  Princeton, NJ 08536 \\
  \texttt{amit.chakraborty@siemens.com}\\
}

\begin{document}

\maketitle

\begin{abstract}
Microstructures of a material form the bridge linking processing conditions - which can be controlled, to the material property - which is the primary interest in engineering applications. Thus a critical task in material design is establishing the processing-structure relationship, which requires domain expertise and techniques that can model the high-dimensional material microstructure. This work proposes a deep learning based approach that models the processing-structure relationship as a conditional image synthesis problem. In particular, we develop an auxiliary classifier Wasserstein GAN with gradient penalty (ACWGAN-GP) to synthesize microstructures under a given processing condition. This approach is free of feature engineering, requires modest domain knowledge and is applicable to a wide range of material systems. We demonstrate this approach using the ultra high carbon steel (UHCS) database, where each microstructure is annotated with a label describing the cooling method it was subjected to. Our results show that ACWGAN-GP can synthesize high-quality multiphase microstructures for a given cooling method.
\end{abstract}

\section{Introduction}
Delineating \emph{processing}–\emph{structure}–\emph{property} relationships constitutes a major focus in design of advanced material systems \citep{olson1997computational}. While analytical and statistical methods have been successfully used for design of certain materials \citep{florescu2009designer, fullwood2010microstructure, lee2017concurrent}, the underlying assumptions on homogeneity and isotropy limit their generalizability and transferability to other material systems. To address these challenges, machine learning and data-driven techniques have piqued interest in the material science community. Microstructure reconstruction, by allowing an effective method to understand the high dimensional microstructure space, plays a critical role in computational material design. Prior work along this line has used deep learning to predict material property from microstructure \citep{cecen2018material,cang2018improving, yang2018microstructural}, reconstruct statistically equivalent microstructures \citep{li2018transfer}, and synthesize microstructures with desired properties \citep{yang2018microstructural, cang2018improving}.

Generative models, such as variational autoencoders (VAE) \citep{kingma2014adam} and generative adversarial networks (GAN) \citep{goodfellow2014generative}, are key enablers of deep learning based microstructure reconstruction. \cite{cang2018improving} used VAEs to synthesize two-phase microstructures and demonstrated that convolutional networks can be used for material property prediction. \cite{yang2018microstructural} used deep convolutional GAN to synthesize microstructures and transfer learning to improve structure-property predictions. Both of these works augmented the generative model loss function with style transfer and mode collapse losses. \cite{singh2018physics} leveraged WGAN-GP and used generative invariance checker \& discriminator concurrently to generate two-phase microstructures.

However, this line of work on using generative models for microstructure reconstruction has two limitations. First, previous works have focused on two-phase microstructures, while many material systems comprise multiphase microstructures. Characterization and reconstruction of multiphase materials have been studied scarcely, especially due to the fact that evaluating higher order correlation for multiphase materials is challenging. To the best of our knowledge, transfer learning technique by \cite{li2018transfer} is the only method that has been used to reconstruct multiphase microstructures. This method uses first few convolutional layers of a pre-trained VGG16 \citep{simonyan2014very} network to minimize difference in Gram Matrices of the target and reconstructed image. Although accurate, this method can only reconstruct images that match a single target microstructure and cannot model a distribution in the way generative models do. Second, previous works do not account for the influence of processing conditions on microstructure. This is a key aspect in material design since we strive to not only design an optimal microstructure but also identify the processing conditions necessary to manufacture it.

We address these two challenges by developing an auxiliary classifier Wasserstein GAN with gradient penalty (ACWGAN-GP) to synthesize multiphase alloy microstructures from user defined processing methods and demonstrate this approach using the Ultra High Carbon Steel Database \citep{decost2017uhcsdb}. Modelling this dataset is extremely challenging owing to the multiphase, heterogeneous microstructures it contains. The key contributions of this work are:
\begin{itemize}[noitemsep,topsep=-10pt]
\item We demonstrate that GANs can synthesize multiphase microstructure images.
\item We demonstrate that ACWGAN-GP enables us to condition the generator on a critical processing condition, namely the cooling method.
\item We use VGG16-based feature extractor and t-SNE to validate the proposed approach.
\end{itemize}

%
\section{Learning Framework}
\subsection{Auxiliary Classifier Wasserstein GAN with Gradient Penalty (ACWGAN-GP)}
%
To learn underlying latent distributions from high-dimensional data, such as images, GANs formulate the learning problem as a two-player zero-sum game between a generator (which tries to generate synthetic images indistinguishable from the real ones) and a discriminator (which tries to distinguish whether an image is real or has been synthesized). Furthermore, by imposing additional structure into the GAN latent space, conditional generative models provide an efficient means to better control features in synthesized samples. Conditional GAN \citep{mirza2014conditional}, by providing side information (e.g. class labels) to both generator and discriminator, proposes an implementation of this approach and subsequently improves visual quality and diversity of the synthesized images. A later work by \cite{odena2017conditional} introduces the auxiliary classifier GAN (AC-GAN) architecture which, in addition to using class labels for synthesizing class conditional image samples, also includes a classifier which predicts class labels for images. This current work expands upon this architecture to synthesize microstructures from given values of processing parameters.

After GAN was introduced by \cite{goodfellow2014generative}, several improvements have been proposed to achieve stable training and fast convergence. Wasserstein GAN \citep{arjovsky2017wasserstein}, by using earth-mover distance (Wasserstein-1 metric) as a geometrically meaningful measure of mismatch between probability distributions, introduced a loss function that correlates with the quality of synthesized images. A later work by \cite{gulrajani2017improved} has shown that penalizing the $L_{2}$-norm of discriminator gradient, in stead of constraining the discriminator weights, provides a better alternative to enforce the Lipschitz constraint. This change in the discriminator training results in faster convergence as well as overall improvement in the quality of synthesized images.

In this work, to synthesize alloy microstructures conditioned on processing methods, we develop a hybrid framework that uses label conditioning suggested in AC-GAN architecture but leverages Wasserstein-1 metric to design loss functions. Additionally, we use gradient penalty instead of weight clipping for discriminator training. In particular, the generator ($\mathcal{G}$) and the critic ($\mathcal{C}$) in this ACWGAN-GP framework tries to minimize the following loss functions
\begin{displaymath}
\Scale[0.97]
{
L_{_{\mathcal{G}}} =
\underbrace{
- \underset{\tilde{\mathbf{x}}\sim\mathbb{P}_G}{\mathbb{E}}\big[D(\tilde{\mathbf{x}})\big]
}_{\text{Generator loss}}
+
\lambda_2
\underbrace{
\underset{\tilde{\mathbf{x}}\sim\mathbb{P}_G}{\mathbb{E}}
\left[-\log\left(\text{Prob}\big[C(\tilde{\mathbf{x}})=c_{\tilde{\mathbf{x}}}\big]\right)\right]
}_{\text{Sparse cross-entropy over synthetic images}}
}
\end{displaymath}
and
\begin{displaymath}
\Scale[0.95]
{
L_{_{\mathcal{C}}} = 
\underbrace{
\underset{\tilde{\mathbf{x}}\sim\mathbb{P}_G}{\mathbb{E}}\big[D(\tilde{\mathbf{x}})\big]
-
\underset{\mathbf{x}\sim\mathbb{P}_r}{\mathbb{E}}\Big[D(\mathbf{x})\Big]
}_{\text{Discriminator loss}}+
\lambda_1
\underbrace{\underset{\hat{\mathbf{x}}\sim\mathbb{P}_{\hat{x}}}{\mathbb{E}}\big[(\|\nabla_{\hat{x}}D({\hat{\mathbf{x}}})\|_2-1)^2\big]
}_{\text{Gradient-penalty}}+
\lambda_2
\underbrace{
{\mathbb{E}}
\left[-\log\left(\text{Prob}\big[C({x})=c_{{x}}\big]\right)\right]
}_{\text{Sparse Cross-entropy over all images}}
},
\end{displaymath}
where, $\lambda_1=10$, $\lambda_2=1$, $c_{{x}}$ is the true label associated with an image (both real and synthesized) $x$, and $\mathbb{P}_r$ and $\mathbb{P}_G$ represent the distributions of real and synthesized images, respectively. Moreover, $\mathbb{P}_{\hat{x}}$ denote the distribution of samples which have been drawn uniformly along straight lines between pairs of images sampled from $\mathbb{P}_r$ and $\mathbb{P}_G$.
%
%
%
\begin{figure}[t!]
\centering
\includegraphics[width=\textwidth]{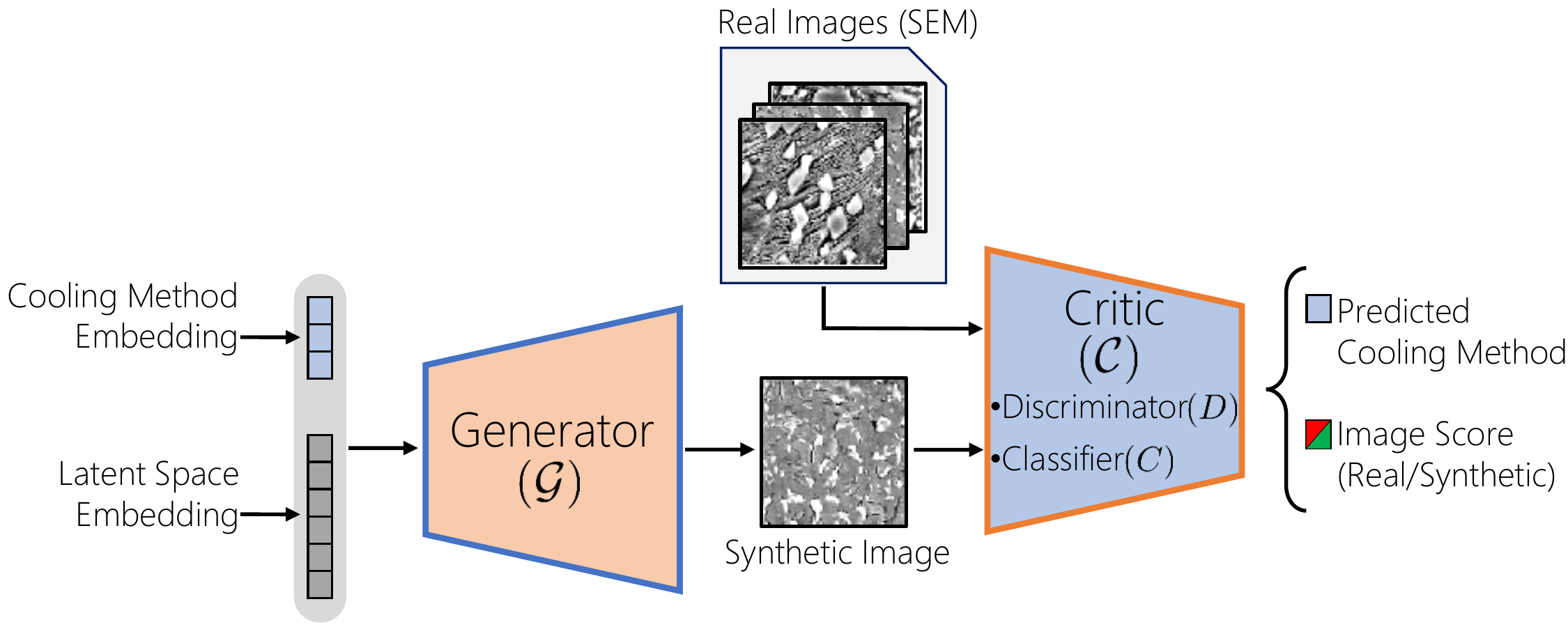}
\vspace{-0.3cm}
\caption{\emph{ACWGAN-GP framework for learning conditional generative models to predict multi-phase microstructures from cooling methods.}}
\label{learning_framework}
\vspace{-0.5cm}
\end{figure}

\subsection{Network architecture and training}
In this work, we represent the cooling method via a 20-dimensional embedding vector and concatenate it with a 100-dimensional Gaussian noise vector. This 120-dimensional combined vector is then supplied to the generator $\mathcal{G}$. The first layer of $\mathcal{G}$ is a Fully connected network with 1024 neurons, followed by a dropout layer with rate = 0.25. This is followed by three Upsampling-Convolution-Leaky ReLU blocks and a final Upsampling-Convolution block with $\tanh$-activation. The critic $\mathcal{C}$ is an approximate mirror image of the generator with four Convolution-Leaky ReLU blocks. These operations extract a 1024-dimensional feature vector which is then passed through a dropout layer with rate = 0.25 and subsequently used by two separate fully connected layers to determine the image score and the corresponding cooling method.

\section{Experiment}
\subsection{The Ultra High Carbon Steel DataBase (UHCSDB)}
UHCSDB is a collection of 961 microstructures obtained from Scanning Electron Microscopy (SEM) of samples with identical composition but subjected to varied heat treatments. The variation in heat treatment influences the microstructure and relevant properties. SEM microstructures curated in UHCSDB correspond to 5 different cooling methods (\emph{no heat treatment}, \emph{quenching}, \emph{furnace cooling}, \emph{air cooling} and \emph{constant heating at $650^{\circ}$C for 1 Hour}). Within this database, we focus on microstructures captured at a magnification of 10.3 pix/micron. To train our ACWGAN-GP model to synthesize microstructures of size 128x128 pixels, we created a dataset containing random crops from the original 172 SEM microstructures in UHCSDB. Also, to mitigate the imbalance in the number of images corresponding to different cooling methods, we adjust the number of random crops in such a way that each of the cooling methods have approx. 1400 images. In addition we augment the dataset by rotating the images by 90, 180 and 270 degrees.
%
%
\begin{figure}[t!]
\centering
\includegraphics[width=\textwidth]{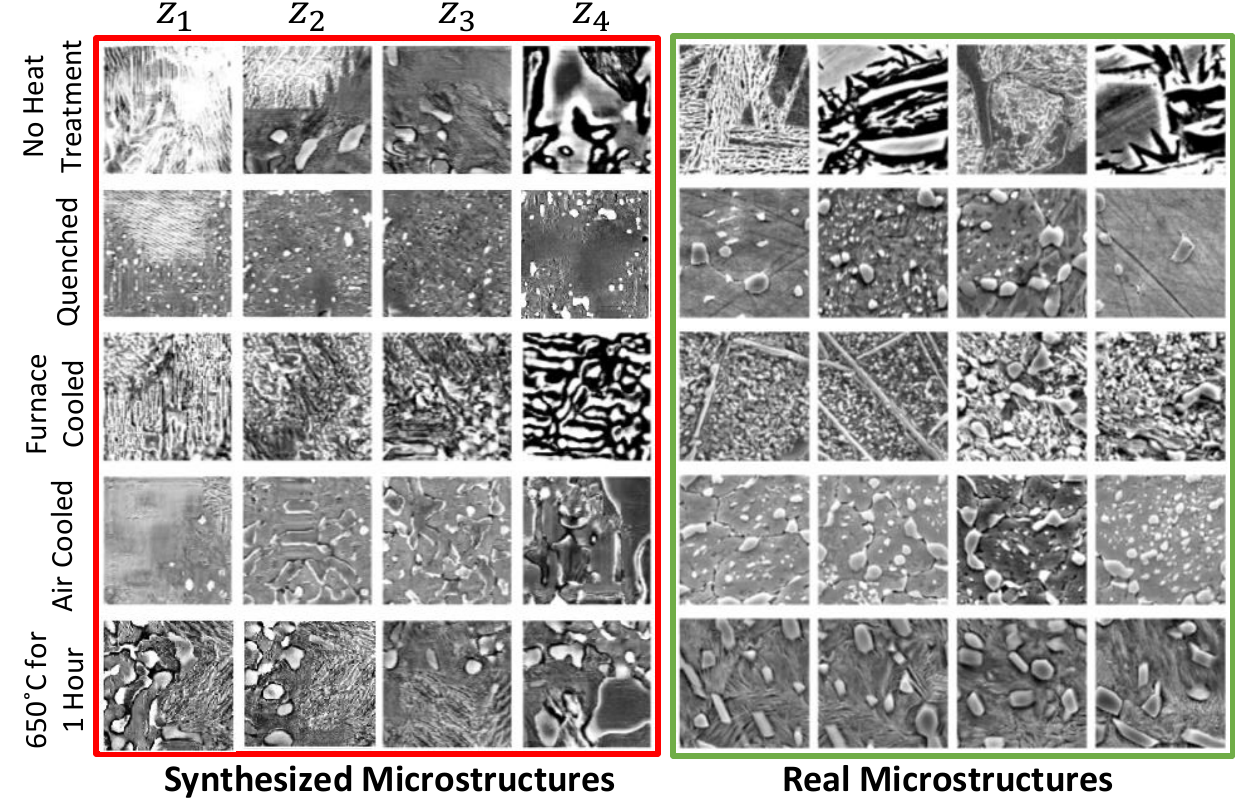}
\vspace{-0.3cm}
\caption{\emph{Visual comparison of synthesized microstructures (after 6000 epochs) with the real ones.}}
\label{Visual_Eval}
\vspace{-0.5cm}
\end{figure}

\subsection{Results}
We train the ACWGAN-GP for 6000 epochs with a batch size of 64 and use Adam Optimizer \citep{kingma2014adam} with a learning rate of $5\times 10^{-5}$, while keeping $\beta_{1}=0.5$ \& $\beta_{2}=0.9$. Also, $\mathcal{C}$:$\mathcal{G}$ training ratio is maintained at 5:1. Figure~\ref{Visual_Eval} shows a few representative images synthesized by ACWGAN-GP along with real microstructures for visual comparison. Each column of synthesized images correspond to the same Gaussian noise vector $z_{i}$ while each row corresponds to a specific cooling method. Although this figure highlights the visual resemblance between real and synthesized microstructures, we present rigorous evaluation methods in what follows.
\begin{figure}[t!]
\centering
\includegraphics[width=\textwidth]{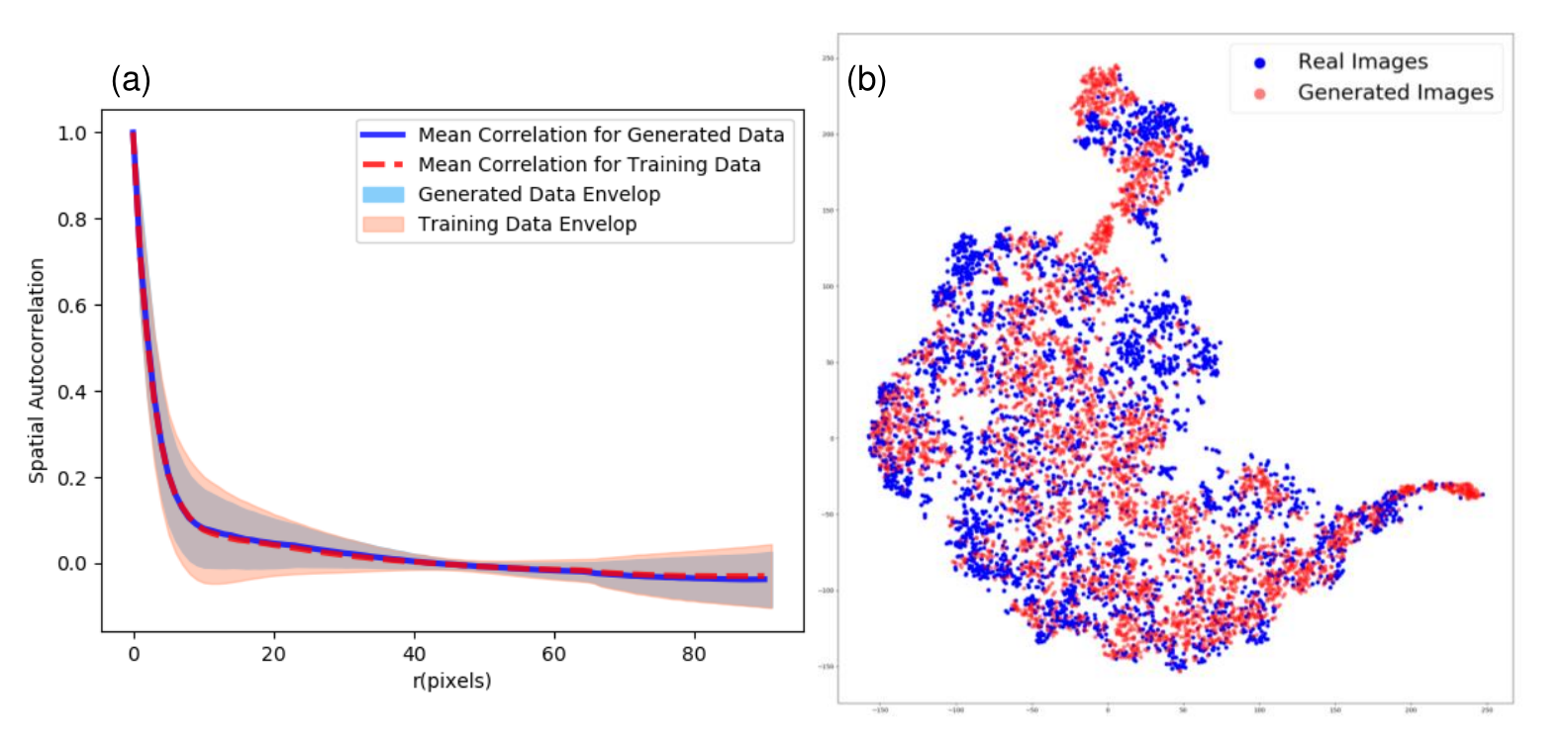}
\vspace{-0.3cm}
\caption{\emph{Validation of the trained ACWGAN-GP model using (a) 2-point spatial correlation and (b) 2-dimensional projection of a t-SNE based embedding associated with real and synthesized microstructures.}}
\label{validation}
\vspace{-0.5cm}
\end{figure}

Furthermore, to perform a quantitative evaluation of the synthesized microstructures, we randomly select 4000 pairs of real and synthesized images and compare their 2-point spatial correlations \citep{yeong1998reconstructing}. Figure~\ref{validation}(a) shows a good match between the correlation values of the real and the synthesized microstructures. However, for complex, multiphase microstructures, this is only a necessary condition, not sufficient. We provide an alternative evaluation which first extracts appropriate feature vectors by using a pre-trained VGG16 and then uses t-SNE based dimensionality reduction to project these vectors onto a 2-dimensional space. We applied this procedure to 4000 pairs of real \& synthesized images to obtain Fig.~\ref{validation}(b). This figure shows vast regions of overlap between real and synthesized microstructures, indicating similarities between them. 
%

\section{Conclusion}
To the best of our knowledge, this is the first attempt to model processing-structure relationship as a conditional image synthesis problem. To accomplish this objective we have utilized the ACWGAN-GP framework, which inherits training stability of WGAN-GP and conditional image generation of ACGAN. We use this framework for reconstructing multiphase microstructures from UHCSDB and demonstrate its capability in synthesizing high quality microstructures from a given cooling method. In our future work, we will introduce additional processing parameters, such as annealing temperature and time, to achieve tighter control and better insight over the synthesized microstructures.

\bibliographystyle{plainnat}
\bibliography{ACWGANGP_AlloyMStrct_arXiv.bib}

\begin{thebibliography}{18}
\providecommand{\natexlab}[1]{#1}
\providecommand{\url}[1]{\texttt{#1}}
\expandafter\ifx\csname urlstyle\endcsname\relax
  \providecommand{\doi}[1]{doi: #1}\else
  \providecommand{\doi}{doi: \begingroup \urlstyle{rm}\Url}\fi

\bibitem[Arjovsky et~al.(2017)Arjovsky, Chintala, and
  Bottou]{arjovsky2017wasserstein}
Martin Arjovsky, Soumith Chintala, and L{\'e}on Bottou.
\newblock Wasserstein {GAN}.
\newblock \emph{arXiv preprint arXiv:1701.07875}, 2017.

\bibitem[Cang et~al.(2018)Cang, Li, Yao, Jiao, and Ren]{cang2018improving}
Ruijin Cang, Hechao Li, Hope Yao, Yang Jiao, and Yi~Ren.
\newblock Improving direct physical properties prediction of heterogeneous
  materials from imaging data via convolutional neural network and a
  morphology-aware generative model.
\newblock \emph{Computational Materials Science}, 150:\penalty0 212--221, 2018.

\bibitem[Cecen et~al.(2018)Cecen, Dai, Yabansu, Kalidindi, and
  Song]{cecen2018material}
Ahmet Cecen, Hanjun Dai, Yuksel~C Yabansu, Surya~R Kalidindi, and Le~Song.
\newblock Material structure-property linkages using three-dimensional
  convolutional neural networks.
\newblock \emph{Acta Materialia}, 146:\penalty0 76--84, 2018.

\bibitem[DeCost et~al.(2017)DeCost, Hecht, Francis, Webler, Picard, and
  Holm]{decost2017uhcsdb}
Brian~L DeCost, Matthew~D Hecht, Toby Francis, Bryan~A Webler, Yoosuf~N Picard,
  and Elizabeth~A Holm.
\newblock {UHCSDB: UltraHigh carbon steel micrograph database}.
\newblock \emph{Integrating Materials and Manufacturing Innovation}, 6\penalty0
  (2):\penalty0 197--205, 2017.

\bibitem[Florescu et~al.(2009)Florescu, Torquato, and
  Steinhardt]{florescu2009designer}
Marian Florescu, Salvatore Torquato, and Paul~J Steinhardt.
\newblock Designer disordered materials with large, complete photonic band
  gaps.
\newblock \emph{Proceedings of the National Academy of Sciences}, 106\penalty0
  (49):\penalty0 20658--20663, 2009.

\bibitem[Fullwood et~al.(2010)Fullwood, Niezgoda, Adams, and
  Kalidindi]{fullwood2010microstructure}
David~T Fullwood, Stephen~R Niezgoda, Brent~L Adams, and Surya~R Kalidindi.
\newblock Microstructure sensitive design for performance optimization.
\newblock \emph{Progress in Materials Science}, 55\penalty0 (6):\penalty0
  477--562, 2010.

\bibitem[Goodfellow et~al.(2014)Goodfellow, Pouget-Abadie, Mirza, Xu,
  Warde-Farley, Ozair, Courville, and Bengio]{goodfellow2014generative}
Ian Goodfellow, Jean Pouget-Abadie, Mehdi Mirza, Bing Xu, David Warde-Farley,
  Sherjil Ozair, Aaron Courville, and Yoshua Bengio.
\newblock Generative adversarial nets.
\newblock In \emph{Advances in Neural Information Processing Systems}, pages
  2672--2680, 2014.

\bibitem[Gulrajani et~al.(2017)Gulrajani, Ahmed, Arjovsky, Dumoulin, and
  Courville]{gulrajani2017improved}
Ishaan Gulrajani, Faruk Ahmed, Martin Arjovsky, Vincent Dumoulin, and Aaron~C
  Courville.
\newblock Improved training of wasserstein gans.
\newblock In \emph{Advances in Neural Information Processing Systems}, pages
  5767--5777, 2017.

\bibitem[Kingma and Ba(2014)]{kingma2014adam}
Diederik~P Kingma and Jimmy Ba.
\newblock Adam: A method for stochastic optimization.
\newblock \emph{arXiv preprint arXiv:1412.6980}, 2014.

\bibitem[Lee et~al.(2017)Lee, Yu, Engel, Reese, Rhee, Chen, and
  Odom]{lee2017concurrent}
Won-Kyu Lee, Shuangcheng Yu, Clifford~J Engel, Thaddeus Reese, Dongjoon Rhee,
  Wei Chen, and Teri~W Odom.
\newblock Concurrent design of quasi-random photonic nanostructures.
\newblock \emph{Proceedings of the National Academy of Sciences}, 114\penalty0
  (33):\penalty0 8734--8739, 2017.

\bibitem[Li et~al.(2018)Li, Zhang, Zhao, Burkhart, Brinson, and
  Chen]{li2018transfer}
Xiaolin Li, Yichi Zhang, He~Zhao, Craig Burkhart, L~Catherine Brinson, and Wei
  Chen.
\newblock A transfer learning approach for microstructure reconstruction and
  structure-property predictions.
\newblock \emph{Nature Scientific Reports}, 8, 2018.

\bibitem[Mirza and Osindero(2014)]{mirza2014conditional}
Mehdi Mirza and Simon Osindero.
\newblock Conditional generative adversarial nets.
\newblock \emph{arXiv preprint arXiv:1411.1784}, 2014.

\bibitem[Odena et~al.(2017)Odena, Olah, and Shlens]{odena2017conditional}
Augustus Odena, Christopher Olah, and Jonathon Shlens.
\newblock {Conditional image synthesis with auxiliary classifier GANs}.
\newblock In \emph{Proceedings of the 34th International Conference on Machine
  Learning}, pages 2642--2651, 2017.

\bibitem[Olson(1997)]{olson1997computational}
Gregory~B Olson.
\newblock Computational design of hierarchically structured materials.
\newblock \emph{Science}, 277\penalty0 (5330):\penalty0 1237--1242, 1997.

\bibitem[Simonyan and Zisserman(2014)]{simonyan2014very}
Karen Simonyan and Andrew Zisserman.
\newblock Very deep convolutional networks for large-scale image recognition.
\newblock \emph{arXiv preprint arXiv:1409.1556}, 2014.

\bibitem[Singh et~al.(2018)Singh, Shah, Pokuri, Sarkar, Ganapathysubramanian,
  and Hegde]{singh2018physics}
Rahul Singh, Viraj Shah, Balaji Pokuri, Soumik Sarkar, Baskar
  Ganapathysubramanian, and Chinmay Hegde.
\newblock Physics-aware deep generative models for creating synthetic
  microstructures.
\newblock \emph{arXiv:1811.09669}, 2018.

\bibitem[Yang et~al.(2018)Yang, Li, Brinson, Choudhary, Chen, and
  Agrawal]{yang2018microstructural}
Zijiang Yang, Xiaolin Li, L~Catherine Brinson, Alok~N Choudhary, Wei Chen, and
  Ankit Agrawal.
\newblock Microstructural materials design via deep adversarial learning
  methodology.
\newblock \emph{Journal of Mechanical Design}, 140\penalty0 (11):\penalty0
  111416, 2018.

\bibitem[Yeong and Torquato(1998)]{yeong1998reconstructing}
CLY Yeong and Salvatore Torquato.
\newblock Reconstructing random media.
\newblock \emph{Physical Review E}, 57\penalty0 (1):\penalty0 495, 1998.

\end{thebibliography}

\end{document}